# New Concept for Electron Beam-Dump Experiment Utilizing Directional WIMP Detector

D.P. Snowden-Ifft[1], J.L. Harton[2], Nan Ma[1], F.G. Schuckman II[2]
[1]*Physics Department, Occidental College, Los Angeles, California 90041*
[2]*Department of Physics, Colorado State University, Fort Collins, CO 80523-1875, U.S.A.*

Light dark matter in the context of dark sector theories is an attractive candidate for the dark matter thought to make up the bulk of the mass of our universe. We explore here the possibility of using a low-pressure, negative-ion, time projection chamber detector to search for light dark matter behind the beam dump of an electron accelerator. The sensitivity of a 10 m long detector is several orders of magnitude better than existing limits. This sensitivity includes regions of parameter space where light dark matter is predicted to have a required relic density consistent with measured dark matter density. Backgrounds at shallow depth will need to be considered carefully. However, several signatures exist, including a powerful directional signature, which will allow a detection even in the presence of backgrounds.

**Introduction**

Despite decades of astounding experimental progress in direct searches for dark matter in the GeV-TeV mass-scale [1], there are no compelling detections to date. This absence of detections, together with the lack of any hint of supersymmetry at the LHC [2] places severe constraints on the minimal, most 'natural', dark matter models. That, in turn, has led both theorists and experimentalists to look beyond the classic, supersymmetry-motivated weakly interacting massive particle (WIMP) dark matter [3][4]. An interesting candidate scale is light dark matter in the range MeV-GeV [5]. Such particles find a natural home in theories which postulate new MeV-GeV scale 'dark' force carriers [5] and are accessible at high intensity accelerators with specially designed detectors [4]. This paper examines the possibility of utilizing a directional WIMP time projection chamber (TPC) [6] to search for light dark matter at accelerators (LDMA).

**Dark Sectors and Light Dark Matter**

Electron beam dump experiments have a history dating back to the 1980s [7]. Recently there has been renewed interest in them because they have been shown to have high sensitivity to light dark matter under the parameterization of dark sector theories [4][8]. A schematic,

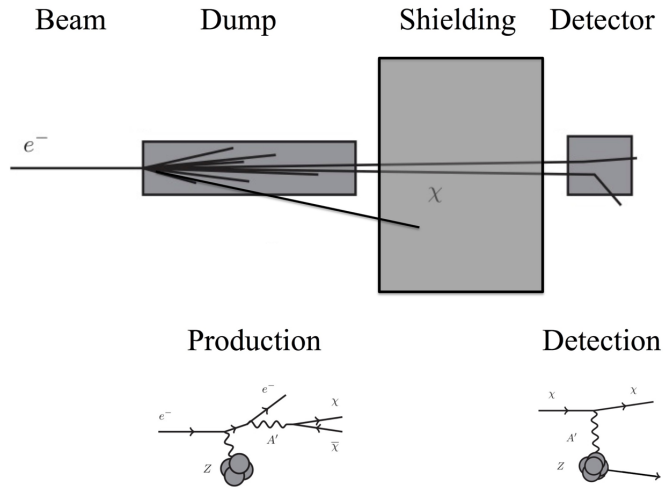

Figure 1 – Schematic showing the major elements of a beam dump experiment.

highlighting the major elements of a beam dump experiment, is shown in Figure 1. The four main elements of a beam dump experiment are a multi-GeV electron beam, an accelerator dump, shielding to stop standard model particles produced in the dump and a detector. Light dark matter particles would be produced when the electron beam interacts with the nuclei in the beam dump via the Cabibbo-Parisi radiative process producing $\chi\bar{\chi}$ pairs [4]. If the mass of the mediator $A'$, $m_{A'}$, is smaller than twice the mass of the dark matter particles, $m_\chi$ $\left(m_{A'} < 2m_\chi\right)$ then the dominant production mechanism is the radiative process illustrated in Figure 1 with $A'$ off-shell [4]. In this regime, the production scales as $\sim \alpha_D \varepsilon^2 / m_\chi^2$ where $\alpha_D$ is the dark sector equivalent to the fine structure constant and $\varepsilon$ governs the coupling strength between the dark sector and the normal electromagnetic sector. Both are related to couplings in the Lagrangian [4]. If

$m_{A'} > 2m_\chi$ then the dominant production mechanism is the radiative production of the $A'$ followed by decay into a $\chi\bar{\chi}$ pair [4], also illustrated in Figure 1 on the left. In this regime, the production scales as $\sim \varepsilon^2 / m_{A'}^2$ [4]. The Beam Dump eXperiment (BDX) has been exploring the sensitivity and capability of a NaI scintillator detector to the dark sector [9][10] through various inelastic channels with a threshold, on shower energy, above ~100 MeV. Because low-pressure, directional TPCs have thresholds, typically, three orders of magnitude smaller, we will only consider the elastic scattering channel in this paper, shown on the right in Figure 1. The differential, elastic scattering cross-section for coherent detection of the dark matter particles is given, to good approximation, by,

$$\frac{d\sigma}{dT} \approx \frac{-8\pi\alpha\alpha_D \varepsilon^2 Z^2 M}{\left(m_{A'}^2 + 2MT\right)^2} \tag{1}$$

where $T$ is the kinetic energy of the recoiling nucleus in the lab frame, $\alpha$ is the fine structure constant, $M$ and $Z$ are the mass and charge of the scattered nucleus [8].

**The Directional Recoil Identification From Tracks (DRIFT) Detector**

WIMP detectors search for ~keV/amu nuclear recoils caused by dark matter [11]. Directional WIMP detectors go a step further and attempt to measure the direction of the recoiling ions to provide a strong signature of WIMP interactions [12]. Low pressure gaseous detectors are preferred for this work as the recoil ranges are then long enough to be measurable [6]. For the past 20 years DRIFT has utilized negative ion drift to limit diffusion in ~40 Torr of gaseous $CS_2$ [13][14]. The reduction of diffusion in all 3 dimensions allows for the preservation of the few-mm-ionization track information [15][16][17][18]. As discussed in these papers DRIFT has the highest resolution in the drift, z, direction, a fact exploited for the directional signature discussed later in this paper.

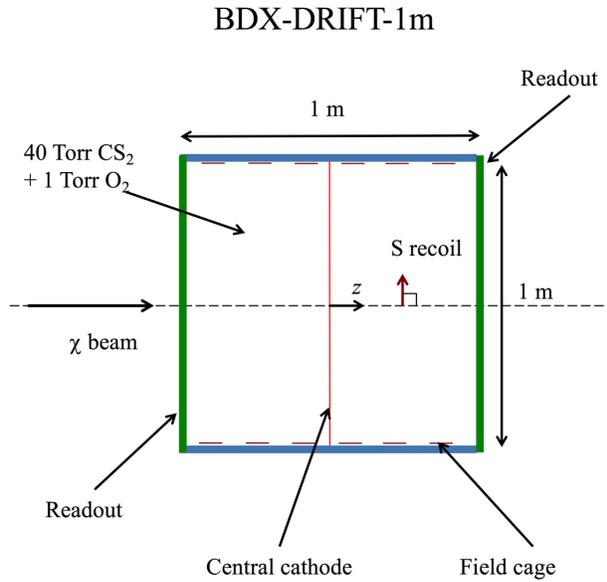

Figure 2 – A sketch of the BDX-DRIFT-1m detector. The lateral, *xy*, dimensions are 1 m each.

We consider a DRIFT-like detector placed behind the beam dump and explore its sensitivity and capabilities for probing the dark sector. A sketch of a BDX-DRIFT-1m module is shown in Figure 2. The accelerator, beam dump and shielding are to the left producing a $\chi$ beam which enters from the left. The readouts on either end couple to two back to back drift volumes filled with a mixture of 40 Torr $CS_2$ and 1 Torr $O_2$ and placed into the beam, as shown. Because of the prevalence of S in the gas and the $Z^2$ dependence for elastic, low-energy scattering, the recoils would be predominantly S nuclei. S recoils with kinetic energies of order a few 10s of keV produced by light dark matter would be scattered within one degree of perpendicular to the beam line due to extremely low-momentum-transfer scattering kinematics. The signatures of light dark matter interactions, therefore, would be a population of events centered on the beamline, with a particular energy distribution and with ionization parallel to the detector readout planes. A BDX-DRIFT-10m detector would be made of 10 such modules aligned along the *z* dimension.

**Sensitivity to the Dark Sector**

For this calculation $N_e = 10^{22}$ electrons on target (EOT) was assumed with an 11 GeV electron beam. For the dark sector parameters, $\alpha_D = 0.5$ and $m_{A'} = 3m_\chi$ were assumed. Dark matter flux numbers were obtained from a detailed Monte Carlo simulation done at INFN Genoa [19] including secondary scattering of the electrons in the dump. The number of detected nuclear recoil scatters was obtained by integrating Equation (1) above $T_{thres} = 20$ keV. Zero background was assumed. Figure 3 shows the sensitivity (ability to exclude at 90% confidence level or greater) of a BDX-DRIFT-10m detector under these assumptions in relation to existing limits and the expectation of dark sector dark matter being a relic from the big bang. As shown in Figure 3 the sensitivity of a BDX-DRIFT-10m detector is significantly better than existing limits and has the potential to probe parameter space favored by a dark sector interpretation of the dark matter relic density.

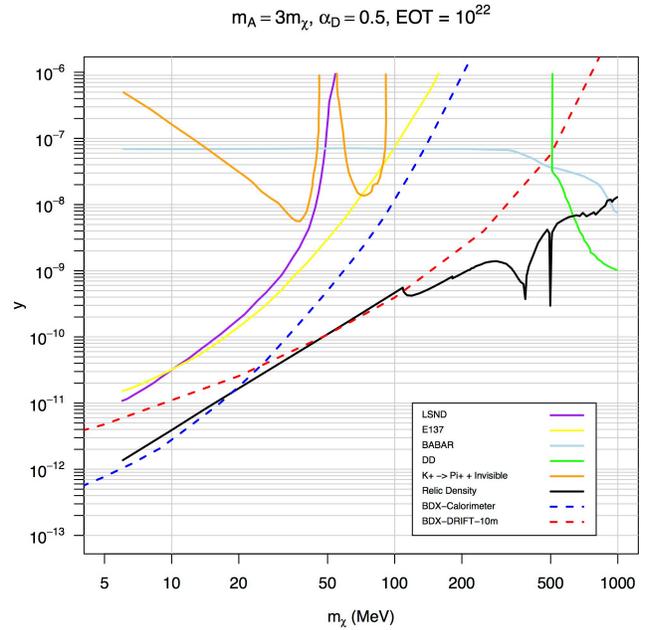

Figure 3 – $m_\chi$ vs $y$ plot exclusion plot. $y = \varepsilon^2 \alpha_D (m_\chi / m_{A'})^4$ is proportional to the annihilation rate allowing for the inclusion of the black thermal relic prediction. 90% confidence level exclusion curves for BDX-DRIFT are shown in dashed red in comparison with other existing limits drawn from [20]. Limits from the companion experiment, BDX-Calorimeter, are shown in dashed blue.

**Backgrounds**

DRIFT has phenomenal gamma rejection due to the difference between high ionization density nuclear recoils and low ionization density Compton electron recoils [21]. For similar reasons DRIFT is insensitive to cosmic ray muons. DRIFT detectors run at ground level [22] have been shown to be sensitive only to nuclear recoils.

In the past DRIFT was background limited by nuclear recoils produced by radon decays on the central cathode [23]. This background was eliminated by the discovery of multiple, ionization-created $CS_2$ anions with introduction of a small amount ~1 Torr of $O_2$ to the $CS_2$ mixture which allowed measurement of the location of the recoil in the drift direction [21][24][25] allowing cathode events to be rejected. DRIFT, running 1 km underground and surrounded by neutron shielding, has been shown to be background free for at least 55 days, producing spin-dependent limits comparable to non-directional solid or liquid based detectors [21][25]. Recoils produced by internal radioactivity are, therefore, well-controlled.

Cosmic ray muons at shallow depth are problematic because they induce neutron emission from nuclei near the detector [26]. Muon induced neutrons are emitted isotropically with energies of ~ 1MeV [26]. We have performed preliminary GEANT [27] simulations of a BDX-DRIFT-10m detector under a 6 m overburden of earth surrounded a 7 mm thick Al vacuum vessel in turn surrounded by 0.75 m of polyethylene shielding. Nuclear recoils (>20 keVr, recoil energy) occur at a rate of ~40 events per day in this volume. Beam currents at existing high-intensity accelerator facilities are of order 100 $\mu A$ requiring, therefore, ~200 days of beam-time to achieve $10^{22}$ EOT with an associated ~8,000 background events. Small signals (~1,000 anions) and long and slow ion drift (10 ms maximum drift time) make it unlikely that timing resolution better than ~10 μs could be achieved [14]. Meanwhile most high-intensity electron accelerators operate with bunch timing several orders of magnitude smaller than this removing timing as an option for cosmic related background suppression. For this reason, a neutron-recoil veto is required.

We consider replacing the polyethylene shielding with either a Gd-loaded water [28] or a Gd-loaded liquid scintillator [29] veto. This will allow for muon tagging of neutron recoil events which simulations indicate account for ~90% of the neutron-recoil background while introducing only 5% deadtime, based on 5 kHz muon rate, from GEANT, and the estimated 10 μs timing resolution. Neutron-recoils not vetoed by muons can still be vetoed. Our GEANT simulations show that over 99.9% of ~MeV energy neutrons producing recoils in the TPC will stop before leaving a veto with about 2% thermalizing in the aluminum walls, but these neutrons can still diffuse to the active veto with high probability. For a DRIFT detector running underground, a Gd-loaded neutron veto was estimated to have >90% efficiency [30]. Efficiency of the veto will depend on photocathode efficiency and coverage on the veto walls, and 10% coverage or better with highly reflective walls has been shown to be effective [28][31]. Detailed simulation work backed by measurements in conjunction with cost and safety considerations will be required to pin down an exact number but we expect >99% veto efficiency based on these initial estimates.

We will also utilize powerful event signatures to allow further background suppression and signal detection.

**Signatures**

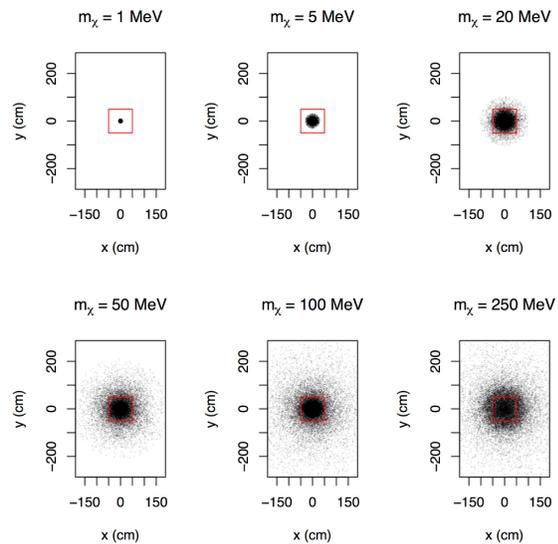

Figure 4 – These plots show the χ beam profile for various assumed dark matter masses at the detector location, shown in red.

$\chi\bar{\chi}$ pairs are produced by decay of the *A'* particle as shown in Figure 1. Assuming the mass of the *A'* particle is much less than the beam energy, the decay will occur in a center of mass (CM) at high velocity with respect to the lab frame. Thus $\chi\bar{\chi}$ pairs will be forward-peaked and because of the proximity of the detector to the beam-dump, the recoil event profile is expected to fall off rapidly from the beamline. For fixed beam energy, the higher the mass of the A' particle the lower the velocity of the CM where the decay into dark matter particles occurs and therefore the less forward peaked they will be. Figure 4 shows simulations of χ production including beam scattering in the beam dump [19] for various dark matter masses and $m_{A'} = 3m_{\chi}$. The detector was assumed to be 10 m from the beam dump. The red boxes show the extent of the detector while the points represent the spread of the beam for various assumed $m_{\chi}$ masses. Thus, a simple measurement of recoil event position will yield a powerful signature of dark matter recoils, enable background suppression and provide information on mass.

The recoil energy spectrum of LDMA interactions is given by equation (1). The response of the detector to neutron recoils generated by Cf-252 has been well modeled, see [21], including position and energy dependent efficiencies. Thus, the response of the detector to a LDMA signal can be accurately modeled and compared to the actual results providing another signature.

The directional signature of LDMA recoils in a BDX-DRIFT detector would be low energy S recoils which start out moving parallel to the readout planes as shown in Figure 2. Naively these events would have zero dispersion in *z* (drift direction) providing BDX-DRIFT with a strong directional signature. However, straggling of recoils at these low energies is significant. Figure 5a shows the result of a SRIM [32] simulation of 1,000 50 keV S recoils oriented, originally, perpendicular to the beam, or *z*, or horizontal direction. The signature, small dispersion in *z*, is degraded by straggling.

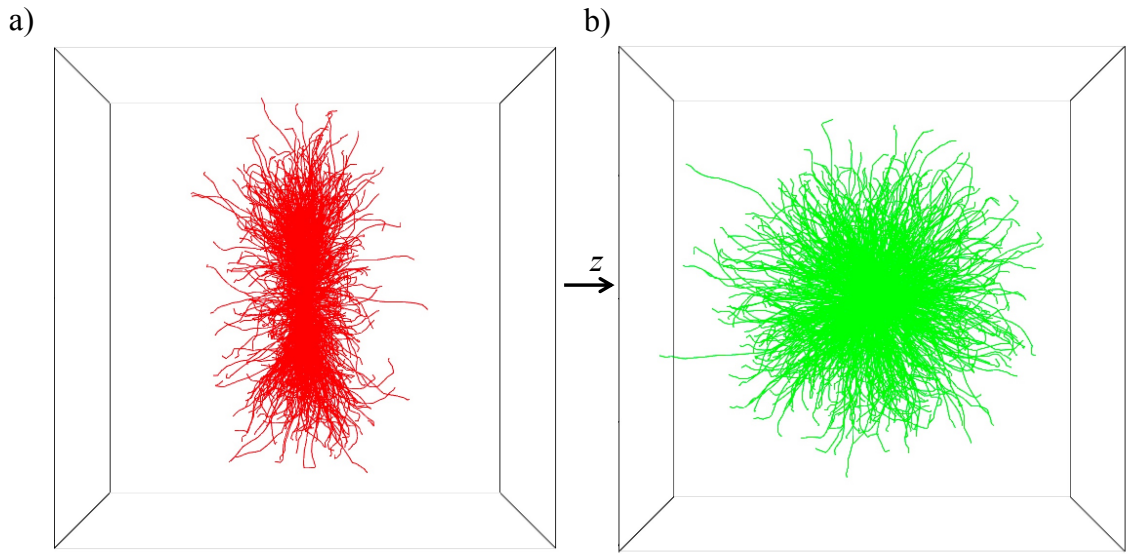

Figure 5a) This figure shows the tracks produced by 1,000 50 keV S recoils originally oriented perpendicular to the beam or z axis according to an SRIM [32] simulation. b) This figure shows 1,000 50 keV S recoils oriented randomly as a comparison background. The surrounding boxes are 4 mm in all dimensions.

For comparison Figure 5b shows a SRIM simulation of 1,000 50 keV S recoils from cosmic ray neutrons. These events are uniformly distributed as expected from the physics of their generation and multiple bounces to enter the fiducial region and confirmed by GEANT simulations. For each event, signal or background, the dispersion of the ionization of the track in $z$, $\sigma_z$, was calculated including diffusion. The distributions are shown in Figure 6. A Kolmogorov-Smirnov (KS) test then determined the probability that $N_s$ signal events with $N_b$ of background events was the same dispersion distribution as $N_s + N_b$ background events. In order to produce a confidence limit (C.L.) this procedure was repeated multiple times with increasing $N_s$ for fixed $N_b$. The number of signal events at which the KS test gave 10% or less probability of similarity 90% of the time was defined to be the 90% C.L. point. The black curves in Figure 7 show the number of signal events, $N_s$, required for a 90% C.L. detection in the presence of $N_b$ background events for three S recoils threshold energies. For zero-background, 16 events would be required at 50 keV recoil energy. But even in the presence of 100 background events, in the area of the detector where signal events are expected, see Figure 4, a significant detection can be found by running the detector only a few times longer than be required for zero background. This is due to the strong directional signature.

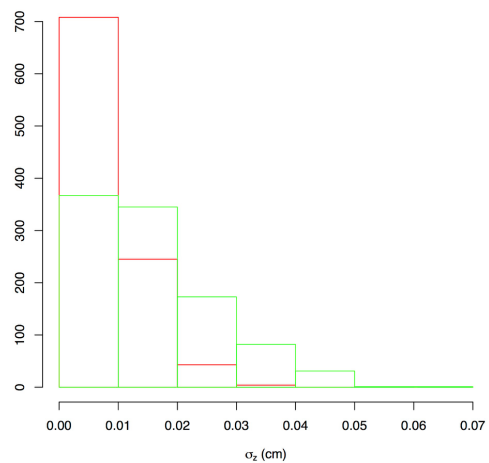

Figure 6 – A histogram showing the difference in the $\sigma_z$ distributions for signal (red) and background (green) events.

Thermal diffusion and various detector effects will contribute to the measured dispersion in *z* as well [14]. The largest of these is thermal diffusion from a track 50 cm from the detector plane. Fortunately, because the absolute position of the event, *z*, can be measured this contribution to the measured dispersion can be subtracted in quadrature [14]. Various detector contributions can also be removed based on [14], though the residual resolution, after subtraction, but has yet to be fully characterized. The green (0.02 cm) and red (0.05 cm) curves in Figure 7 show the effect of adding unaccounted, residual dispersion to the theoretical data.

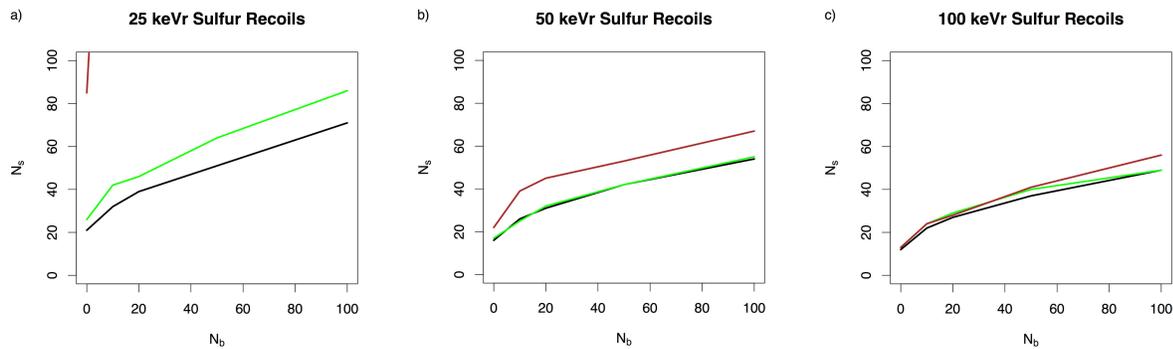

Figure 7 – The figures above show the number of signal events, $N_s$, on the vertical axis required for a 90% C.L. detection in the presence of, $N_b$, background events for three different recoil energies. The black curves are for perfect detector residual resolution, see text. The green curves are for a residual resolution of 0.02 cm. And the red curves are for a residual resolution of 0.05 cm.

**Conclusion**

This paper has explored the possibility of utilizing a low-pressure, negative-ion TPC to search for light dark matter at electron accelerators. A 10 m long detector would have sensitivity significantly better than existing limits and begin to probe the relic density region of parameter space. A veto will be required to mitigate muon-induced neutron-recoils and extensive work will be required to optimize it. Even in the presence of residual background, such detectors can utilize a number of powerful signatures in order to discover signals. Finally discovering a new, non-standard-model particle will not mean that the dark matter search is over. That will require other direct searches. But any hint beyond the standard model will surely move us closer to solving this decades-long problem.

**Acknowledgements**

We would like to thank M. Battaglieri, E. Smith, A. Celentano and S. Vahsen for helpful comments.